\documentclass[twocolumn,prl,aps,epsf,showpacs]{revtex4-1}

\usepackage{graphicx}
\usepackage{dcolumn}
\usepackage{bm}
\usepackage{color}
\usepackage{hyperref}
\begin{document}

\title{Second Landau Level Fractional Quantum Hall Effects in the Corbino Geometry}

\author{B. A. Schmidt$^{1}$, K. Bennaceur$^{1}$, S. Bilodeau$^{1}$, G. Gervais$^{1}$, L. N. Pfeiffer$^{2}$ and K. W. West$^{2}$}

\affiliation{$^{1}$ Department of Physics, McGill University,
Montreal, H3A 2T8, CANADA}

\affiliation{$^{2}$ Department of Electrical Engineering, Princeton University, Princeton NJ 08544 USA}

\date{\today }

\begin{abstract}
For certain measurements, the Corbino geometry has a distinct advantage over the Hall and van der Pauw geometries, in that it provides a direct probe of the bulk 2DEG without complications due to edge effects. This may be important in enabling detection of the non-Abelian entropy of the 5/2 fractional quantum Hall state via bulk thermodynamic measurements. We report the successful fabrication and measurement of a Corbino-geometry sample in an ultra-high mobility GaAs heterostructure, with a focus on transport in the second and higher Landau levels. In particular, we report activation energy gaps of fractional quantum Hall states, with all edge effects ruled out, and extrapolate $\sigma_0$ from the Arrhenius fits. Our results show that activated transport in the second Landau level remains poorly understood. The development of this Corbino device opens the possibility to study the bulk of the 5/2 state using techniques not possible in other geometries.

\end{abstract}

 \maketitle

%TODO: improve LL broadening/ lifetime discussion, quantum vs. classical lifetime?

Ultra-high mobility GaAs/AlGaAs heterostructures subjected to low temperatures and high magnetic fields exhibit a rich set of exotic states, and provide an ideal setting to study the many-body physics of electrons in two dimensions. Owing to the extremely high mobility achievable in these heterostructures, it is possible to observe not only the integer quantum Hall effect (IQHE) and conventional fractional quantum Hall effect (FQHE) \cite{Klitzing, StormerTsui}, but also more exotic and poorly understood FQH states such as the 5/2 and 12/5 FQHE, which are believed theoretically to host non-Abelian quasiparticles.\\

In a quantum Hall state, the bulk of the sample becomes an insulator while the edges host a set of current carrying one-dimensional  channels. This greatly complicates attempts to probe its thermodynamic properties, since electrical and thermal currents in the sample will be primarily sensitive to the physics of the edge. Recent experimental studies of neutral edge modes and tunnelling \cite{Gurman14Nat}, as well as theoretical treatments of edge reconstruction, confirm that the details of transport at QH edges are as-yet poorly understood. One way to avoid the complications of the edge is by eschewing the standard Hall and van der Pauw geometries in favor of a Corbino geometry, in which the electrical and thermal currents between the inner and outer contact are forced to cross (and thus probe) the bulk of the sample.\\

The idea of using a Corbino geometry to avoid the QH edge is not new - indeed it dates back to the earliest thought experiments by Laughlin \cite{Laughlin}. A number of experimental studies have been performed since then, mainly focused on direct measurements of $\sigma_{xx}$ and contactless probing of the sample capacitance and resistance \cite{Dolgopolov92PRB, Jeanneret95PRB, Wiegers99PRB, Khrapai07PRL, Khrapai08PRL} and high frequency effects \cite{Hohls01PRL, Zudov03PRL}. Along those lines, Tomadin et. al. have recently proposed the use of a Corbino 2DEG as an electronic viscometer, analogous to those used to study superfluidity in helium \cite {Tomadin14PRL}. In another direction, thermopower measurements in the Corbino geometry have been proposed as a technique to detect the non-Abelian entropy of the 5/2 FQH state \cite{Barlas12PRB, Ambrumenil13PRL}, and Kobayakawa et. al. \cite{Kobayakawa13JPSP} have reported successful thermopower measurements at around 1~K in a relatively low-mobility device. Similar measurements in the low temperature, high-mobility limit could lead to detection of the predicted non-Abelian entropy of the FQHE. We note that related experimental proposals, such as adiabatic cooling \cite{Gervais10PRL}, would similarly have their interpretation greatly simplified by the use of the Corbino geometry. \\

In this paper, we report transport measurements of an ultra-high mobility GaAs Corbino sample in the quantum Hall regime. Recently, a thorough study of the depinning transition in bubble phases between $\nu = 4$ and $\nu = 5$ by Wang et. al. \cite{Wang14arxiv} was also performed in a high-quality Corbino device. Our focus here is to characterize our sample throughout the FQH regime encompassing much of the second Landau level (SLL),  including at $\nu = 5/2$, and measure the truly bulk activation behaviour of FQH states in ultra-high mobility Corbino samples.\\

The Corbino device was fabricated from a GaAs/AlGaAs heterostructure with quantum well width 30~nm, and with an initial density $\rm 3.06 \times 10^{11} cm^{-2}$ and wafer mobility ${\rm 2.5 \times 10^{7} cm^2/V \cdot s}$ measured at 0.3~K. The zero-field mobility of the sample itself could not be confirmed in the usual way due to the intrinsically two-terminal nature of Corbino measurements, however its magnetotransport behaviour (summarized below) suggests that it was not degraded by the fabrication process. In order to create the Corbino device, contacts were patterned using optical lithography with a Shipley s1813 resist followed by an evaporation (e-beam) of Ge/Au/Ni/Au in 26/54/14/100~nm layers. The contacts were then annealed at 420$^{\circ}$C for 80~s in an $\rm H_2N_2$ atmosphere. The center contact radius was 0.25~mm, and the outer contact had inner and outer radii of 1~mm and 1.5~mm respectively. These contacts were indium soldered to thermally-anchored leads. The sample was mounted on a piece of gold plated silicon wafer (intended for use as a back gate), which was itself mounted on a thermally anchored silver plate. A red LED was used to illuminate the sample during the cooldown, and switched off at 6~K. Data was taken in a dilution refrigerator with base temperature  17~mK using standard lock-in techniques at 13.5~Hz and maximum 50~$\rm \mu V$ excitation, unless otherwise specified.\\

\begin{figure}[h]
\includegraphics[width=1.0\linewidth,angle=0,clip]{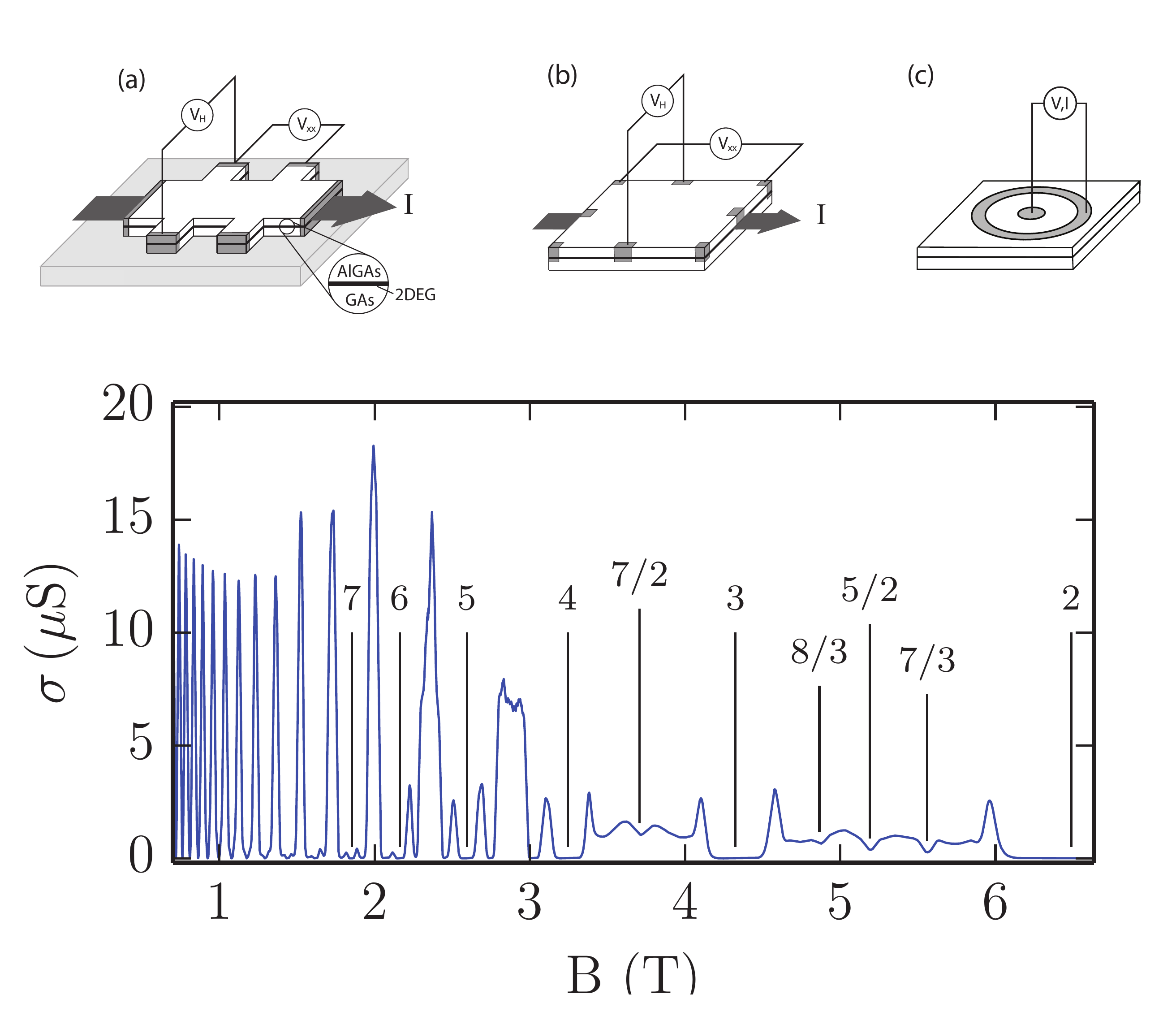}
\caption { Upper panels:  typical transport measurement geometry a) Hall bar, b) van der Pauw and c) Corbino. Bottom panel: conductivity $\sigma_{xx}$ in the Corbino device calculated from the measured conductance as a function of the magnetic field.  }
\label{fig1}
\end{figure}

The Corbino geometry provides direct access to the conductivity via the simple geometric relation 

\begin{equation}
2\pi \sigma_{xx} = (I/V) \ln (r_2 / r_1),
\end{equation}

where $r_1 = 250~\mu$m and $r_2 = 1000~\mu$m in our sample. The usual longitudinal resistivity $\rho_{xx}$ and the conductivity $\sigma_{xx}$ are related via the relation 

\begin{equation}
\rho_{xx} = \frac{\sigma_{xx}}{\sigma_{xx}^2 + \sigma_{xy}^2}, 
\end{equation}

where $\sigma_{xy} = 1/\rho_{xy}$ is the Hall conductivity. As a consequence, the \emph{conductance}, rather than the resistance, vanishes in quantum Hall states measured in the Corbino geometry. Figure~\ref{fig1} shows the conductivity of the sample vs magnetic field up to 7~T. Clear conductivity minima occur at integer and several fractional filling factors, including the 5/2, 7/2 and 7/3 FQH in the SLL. The positions of the lines indicating the filling factors were calculated for a density of $\rm 3.13 \times 10^{11} cm^{-2}$, which was determined based on the position of the 5/2 minimum and agrees with the low-field Shubnikov de Haas oscillations. A similar procedure was used to determine the density for the data in the other figures, which were from different cooldowns or state preparations and thus have slightly different densities. Within a single experimental run, we find that the locations of the robust minima in the IQHE and FQHE series always fit well to a single density, which allows us to confidently identify particular states. \\

\begin{figure}[h]
\includegraphics[width=1.0\linewidth,angle=0,clip]{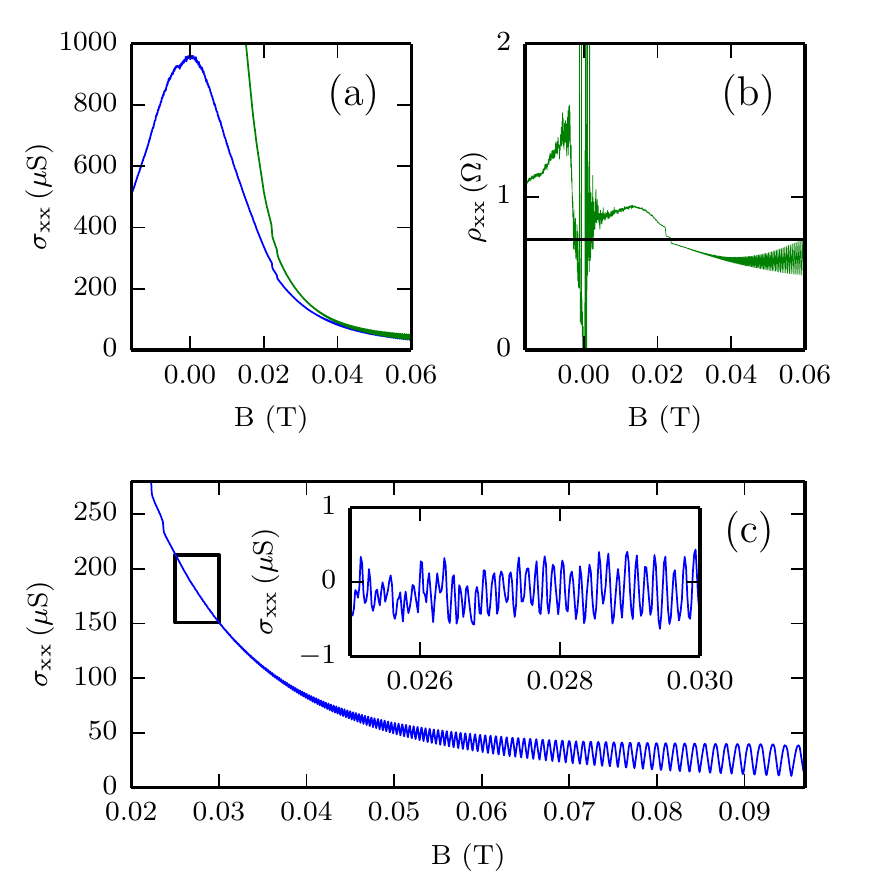}
\caption {(a) Conductivity at low magnetic field calculated directly using equation (1) (blue), and using equation (1) after correcting for a constant lead + contact resistance (green). (b) Resistivity calculated from the measured conductance in conjunction with an assumed classical Hall contribution and fixed lead + contact resistance. The two kinks near 0.02~T are most likely due to superconducting transitions in the sample wiring. The black line indicates the zero-field resistivity corresponding to mobility $\mu = {\rm 2.5 \times 10^{7} cm^2/Vs}$. (c) Shubnikov-de Haas oscillations at low field, measured with a fixed 10~nA current. Inset: SdH oscillations near the extinction point with a smooth background subtracted.}
\label{fig2}
\end{figure}

At zero magnetic field, the resistance of the sample is much smaller than the combined contact and lead resistance, preventing an accurate measurement of the zero-field mobility. Figure~\ref{fig2}(a) shows the conductivity calculated directly from the measurements (blue) and the conductivity with the zero-field contact and lead resistance (totalling 230~$\Omega$) removed (green). For fields above about 0.04~T, the contribution of the contact and lead resistances can be neglected. In order to estimate the zero-field mobility, we have reconstructed the resistivity tensor component $\rho_{xx}$ at low field using some simple approximations. To do so, we have assumed that the Hall resistance at low field is given by the classical linear Hall resistance, with the Hall coefficient determined by the experimentally measured density of the sample.  This is to be expected in high-mobility samples with fixed density, in the low field regime where the 2DEG acts as a Fermi liquid. Using this classical model of $\sigma_{xy}$ along with $\sigma_{xx}$ (corrected for the lead and contact resistance) in equation 2, we calculated $\rho_{xx}$ and the results as shown in Fig.~\ref{fig2}(b). At very low magnetic field, the calculated $\rho_{xx}$ shows a noisy trend due to the dominant lead and contact resistances. However, for magnetic fields above $\approx$ 0.03-0.04~T, our calculated value for $\rho_{xx}$  becomes close to the value expected based on the zero-field mobility previously measured for the same wafer, and indicated  in the figure by a black horizontal line. This gives us confidence that the sample mobility post-fabrication is between 2.1~and 2.5~$\rm \times 10^7 cm^2/V\cdot s$. \\

In order to further characterize the mobility of the sample, we measured Shubnikov-de Haas (SdH) oscillations at low field and extract the quantum lifetime in two different ways \cite{Harrang85PRB}. We can estimate the Landau level broadening from the extinction point of the SdH oscillations, based on the assumption that they only become observable once the cyclotron gap becomes larger than the LL width. Figure~2 shows the SdH oscillations down to their extinction point at roughly 0.026~T, corresponding to a cyclotron gap of 0.52~K and quantum lifetime $\tau_q = 7.3$~ps. The LL width may also be extracted from a Dingle plot (not shown) of the SdH oscillation envelope using $\rho_{xx}$ as calculated for figure 2b. Note that a direct use of $\sigma_{xx}$ leads to a non-linear Dingle plot, due to the contribution of $\sigma_{xy}$ in equation 2 and the contact and lead resistances of the two-point measurement. The resulting (corrected) Dingle temperature is $0.78\pm0.02$~K, corresponding to a quantum lifetime $\tau_q=4.9\pm 0.1$~ps. Both of these estimates are much smaller than the transport lifetime calculated from the sample mobility, $\tau_t = 950$~ps, as expected for samples in which long-range scattering from remote impurities is the primary scattering mechanism. The quantum lifetime is also smaller by about a factor of five than the value  of $\tau = 33  \pm 3$~ps, found for another sample previously measured in our lab \cite{Dean09Thesis}.  We tentatively attribute the lower quantum lifetime to the density, structure of the wells and the location of the dopants (i.e. setback distance). The sample in the current study has roughly twice the density ($\rm 3.06 \times 10^{11} cm^{-2}$ vs. $\rm 1.6 \times 10^{11} cm^{-2}$) and half the setback distance (80~nm vs. 160~nm) of the previously studied one. This may lead to an increase in the scattering rate, and consequently to an increased Landau level width. 

\begin{figure}[h]
\includegraphics[width=1.0\linewidth,angle=0,clip]{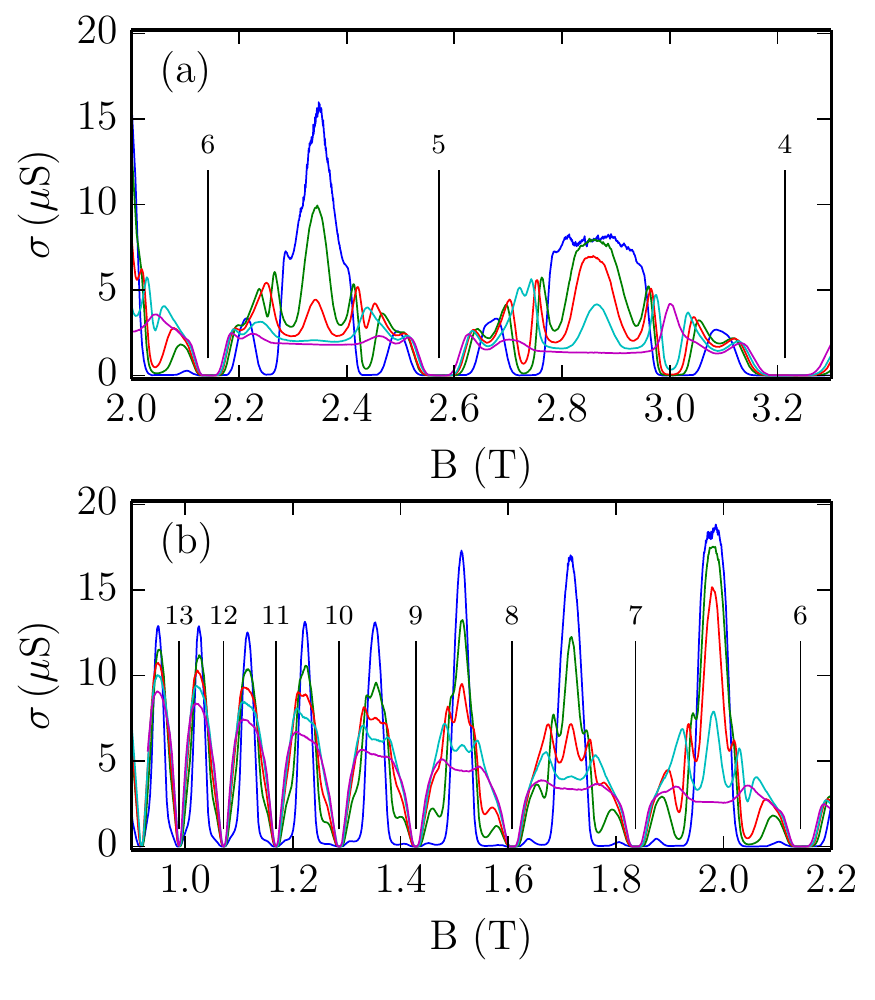}
\caption {Temperature evolution of IQHE and RIQHE in high filling factors at base temperature of 22~mK (blue), 80~mK (green), 100~mK (red), 120~mK (cyan) and 153~mK (purple).  }
\label{fig3}
\end{figure}

Our sample shows remarkably rich behaviour in Landau levels from fillings $\nu=4$ to $\nu=14$, as shown in Fig.~\ref{fig3}.  The well-studied series of states between $\rm \nu = 4, 5$ and 6 are shown in Fig.~\ref{fig3}(a). The anisotropy of the stripe phases, with $\rho_{xx}$ divergent and $\rho_{yy}$ vanishing, is not measurable in the radially symmetric Corbino geometry. Instead, the current flows primarily along the easy direction, leading to a peak in conductance. Since current only flows along one direction, it is effectively confined to an area with the same width as the central contact radius. Particularly in samples with a large ratio $r_2/r_1$, the local current density is much higher than in isotropic states and as a consequence joule heating becomes problematic even for a relatively low current of a few nanoamperes. Thus, the peak of the $\nu = 4  + 1/2$ stripe phase is suppressed in Fig.~\ref{fig3}(a) due to a small DC bias of up to 5~nA driven into the sample by the current preamplifier used for the measurements. However, we have  observed a well-developed peak at $\nu = 4  + 1/2$ on a later cooldown by using a smaller excitation current below 0.4~nA while carefully nulling the DC offset (not shown).\\

As well as these well-known bubble and stripe phases, similar features are shown in Fig.~\ref{fig3}b in even higher Landau levels. Additional reentrant bubble phases are well-developed up to the flanks of $\nu = 9$ and, at base temperature, there are hints of similar states developing up to $\nu = 13$. The stripe phase at half-filling is more difficult to definitively identify, since its hallmark anisotropy cannot be detected and only peaks can be seen.  These are readily observable at $\rm \nu = 6 + 1/2, 7 + 1/2$ and $\rm 8 + 1/2$, and can be distinguished in the 80~mK and 100~mK curves at  $\rm 9 + 1/2$ and  $\rm 10 + 1/2$. The persistence of these features to such high filling factors attests to the extremely high quality of the sample.\\

\begin{figure}[h]
\includegraphics[width=1.0\linewidth,angle=0,clip]{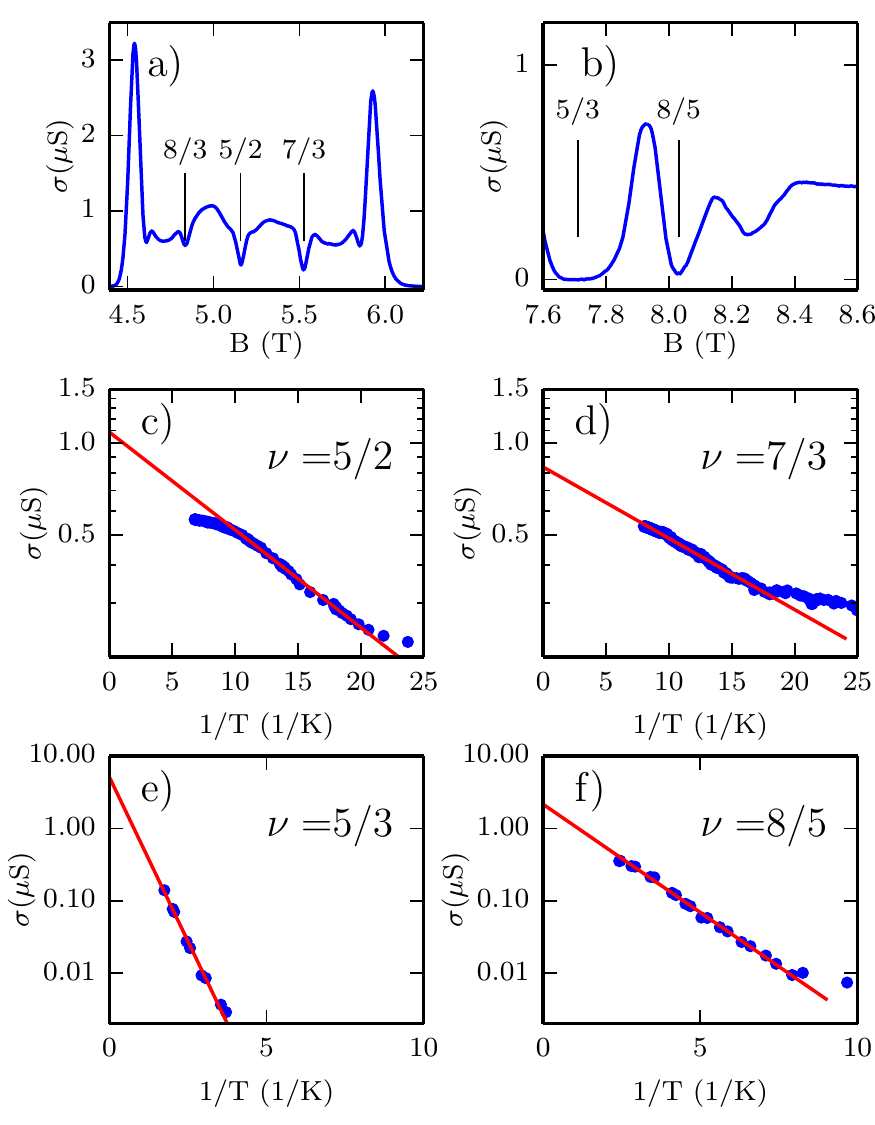}
\caption {a) Magnetoconductance in the SLL at base temperature, $T_{mc} = 23$~mK  showing the 5/2 and b) the 5/3 and 8/5 FQHE. c-f) Arrhenius plot of conductivity minimum versus the inverse temperature at selected filling factors. The red lines are a linear fit to subsets of the data and correspond to the gap energies. These, and the intercept values are reported in Table 1.}
\label{fig4}
\end{figure}

Perhaps the most important feature in the observed magnetoconductance is the presence of fractional quantum Hall states, including the $\nu = 5/2$ in the second Landau level as shown in Fig.~\ref{fig4}(a).  The energy gap of the 5/2 state was measured using the relation $\sigma_{xx}= \sigma_0 e^{- \Delta / 2 k_b T}$, and determined in our device to be  $147 \pm 10$~mK. We tentatively attribute the relatively small gap (compared to Ref. \cite{Nuebler10PRB} and references therein) to poor optimization of the LED lighting condition during the cooling process, which is known to affect features in the second Landau level even beyond its effect on the sample density and mobility \cite{Gamez13PRB}. This measurement is, to our knowledge, the first determination of the $\nu = 5/2$ energy gap in a purely bulk measurement, with any possible edge effects ruled out. We find no evidence for a substantial disagreement between our measurement and conventional van der Pauw gap measurements, suggesting that the latter also reflect the bulk gap rather than edge physics.\\

Since the measurements are performed in the Corbino geometry, we are also able to extract the prefactor $\sigma_0$ from the Arrhenius fit. Theoretical models in the case of a long-range random potential predict $\sigma_0 = 2 (q e)^2 /h$, where $qe$ is the (fractional) quasiparticle or quasihole charge \cite{Polyakov95PRL, Ambrumenil11PRL}. Table~1 shows fit parameters $\Delta$, $\sigma_0$, and $\sigma_0$ in units of $(q e)^2 /h$, for selected FQHE states studied in this work. For the large gap states at $\nu = 5/3$ and $\nu = 8/5$, the experimental estimate of $\sigma_0$ approaches the theoretical prediction. This is consistent with previous experiments \cite{Clark88PRL, Katayama94PRL, Dorozhkin96JETP}, where $\sigma_0$ was found to be closer to $(q e)^2 /h$, and the reduction has been explained by d'Ambrumenil {\it et al.} as an effect of tunnelling \cite{Ambrumenil11PRL}. For the much weaker $5/2$ and $7/3$ states in the SLL, $\sigma_0$ is significantly lower than its predicted value, even with tunnelling taken into account. The measurement of a robust energy gap, even though the extrapolated conductivity is much lower than expected, suggests that transport in this regime is still poorly understood.\\

\begin{table}
	\caption {Arrhenius fit parameters for data shown in Fig.~\ref{fig4}.}
	%\begin{tabular}
	\begin{tabular}{c c c c c}
	$\nu$ &  $\Delta$ [mK] 	& q 		&$\sigma_{0}$ [$\mu$S]	& 	$\sigma_{0} \, [(qe)^2/h]$ \\
	\hline	
	5/2 & $147\pm7$  		& 1/4    	& $1.09 \pm 0.07$	&$0.12\pm0.01$\\
	7/3 & $107\pm5$		& 1/3  	& $0.84 \pm 0.05$	&$0.19\pm0.01$\\
	5/3 & $4180\pm210$		& 1/3		& $5.1 \pm 0.2$	&$1.19\pm0.35$\\
	8/5 & $1380\pm70$		& 1/5 	& $2.2 \pm 0.4$ 	&$1.39\pm0.25$\\
	\end{tabular}
	%\end{ruledtabular}
\end{table}

In conclusion, we have successfully fabricated a Corbino sample with an ultra-high electron mobility and measured its transport properties. Magnetotransport measurements show that it is of similar quality to van der Pauw samples made from similar wafers, {\it i.e.} the mobility has not been significantly degraded by the fabrication process. In particular, our sample shows a series of re-entrant stripe and bubble phase at very high Landau level fillings, and a deep 5/2 fractional quantum Hall minimum in the second Landau Level. Such Corbino devices allow one to probe the bulk of delicate FQH states, without complicating edge effects. In the case of energy gap measurements, we find that the gaps are consistent with those measured in conventional geometries, suggesting that the edge does not play a significant role in those measurements. However, for other measurements, a true bulk probe represents a significant advance. In particular, this work opens the door to thermodynamic measurements of the ground state entropy of possible non-Abelian quantum Hall states such as the 5/2 and 12/5 FQHE.\\

This work has been supported by NSERC, CIFAR and FQRNT. The work at Princeton was partially funded by the Gordon and Betty Moore Foundation as well as the National Science Foundation MRSEC Program through the Princeton Center for Complex Materials (DMR-0819860). Sample fabrication was carried out at the McGill Nanotools Microfabrication facility. We gratefully thank R. Talbot, R. Gagnon, and J. Smeros for technical assistance, and S. Gaucher for assistance with preparation of the manuscript. All data, analysis details and material recipes presented in this work are available upon request to G.G.\\

\end{document}